\documentclass[a4paper,english,conference]{IEEEtran}
\usepackage{latexsym}
\usepackage{float}
\usepackage{amsfonts}
\usepackage{amsbsy}
\usepackage{amssymb}
\usepackage{times}
\usepackage{graphicx}
\usepackage{setspace}
\usepackage{enumerate}
\usepackage[usenames]{color}
\usepackage[dvips]{pstcol}
\usepackage{epstopdf}
\usepackage{cite}
\usepackage{amssymb}
\usepackage{amsfonts}
\usepackage{graphicx}
\usepackage{epsfig}
\usepackage{psfrag}
\usepackage{xcolor}
\usepackage{amsfonts, bm}
\usepackage{epstopdf}
\usepackage{cite}
\usepackage{color}
\usepackage{xcolor}
\usepackage{subfig}
\usepackage{verbatim}
\usepackage{multirow}
\usepackage{array}
\usepackage{booktabs}
\usepackage{amsthm}
\usepackage{makecell}
\usepackage{units}
\usepackage[linesnumbered, ruled]{algorithm2e}
\usepackage{algpseudocode}
\usepackage[ruled]{algorithm2e}
\usepackage{amsmath}

\IEEEoverridecommandlockouts
\begin{document}
\title{Location-Agnostic Channel Knowledge Map Construction for Dynamic Scenes}
\author{\IEEEauthorblockN{Kequan Zhou, Guangyi Zhang, Hanlei Li, Yunlong Cai, and Guanding Yu}
	\IEEEauthorblockA{College of Information Science and Electronic Engineering, Zhejiang University, Hangzhou, China \\ E-mail: \{kqzhou, zhangguangyi, hanleili, ylcai, yuguanding\}@zju.edu.cn}
	\thanks{This work was supported in part by the National Natural Science Foundation of China under Grant 62571477, and in part by Zhejiang Provincial Key Laboratory of Multi-Modal Communication Networks and Intelligent Information Processing, Hangzhou 310027, China.
		
	Our dataset is available at https://github.com/kqzzzz/LAD-CKM-dataset.}
}

\maketitle
\vspace{-3.3em}
\begin{abstract}
To alleviate the pilot and CSI-feedback burden in 6G, channel knowledge map (CKM) has emerged as a promising approach that predicts CSI solely from user locations.
Nevertheless, accurate location information is rarely available in current systems.
Moreover, the uncertainty inherent to highly dynamic scenes further degrades the performance of existing schemes that typically assume quasi-static scenarios.
In this paper, we propose a novel framework named location-agnostic dynamic CKM (LAD-CKM).
Specifically, LAD-CKM is constructed through dynamic radio frequency (RF) radiance field rendering, which takes instantaneous uplink CSI and partial downlink CSI as inputs.
To enable effective rendering, a dedicated radiator representation network (RARE-Net) is designed to capture the spatial-spectral correlations within the inputs.
Furthermore, an adaptive deformation module is devised to deform the uplink CSI-based queries of RARE-Net according to instantaneous channel dynamics, thereby enhancing CSI prediction accuracy under mobility.
In addition, a novel synthetic channel dataset is created in outdoor dynamic scenes via ray-tracing.
Simulation results demonstrate that LAD-CKM yields significant performance gains compared with existing baselines in terms of effective data rate.
\end{abstract}

\IEEEpeerreviewmaketitle

\section{Introduction}
With 6G communications expected to embrace massive multiple-input multiple-output (MIMO) and ultra-wide bandwidths \cite{Wang2021a}, the resulting overhead of conventional pilot-based CSI acquisition approaches becomes prohibitive.
To mitigate this, channel knowledge map (CKM) has recently emerged as a promising alternative \cite{Zeng2024a, Liu2025channel, Wang2025towards}, which enables direct CSI prediction from user locations, thereby significantly reducing CSI acquisition overhead.
Despite its potential to enhance future 6G networks, the effective construction of CKM remains a critical challenge requiring further investigation.

Existing CKM construction methods can generally be divided into two paradigms: location-based and location-agnostic approaches, distinguished primarily by their reliance on explicit user location inputs.
Location-based approaches adhere to the classical CKM definition that establishes a bidirectional mapping between user location and CSI \cite{Xie2019md}.
Specifically, the authors in \cite{Wu2024environment} proposed channel angle map (CAM) and beam index map (BIM) to learn multipath angular parameters and optimal beam indices, respectively, from user locations, resulting in reduced pilot overheads.
In \cite{Orekondy2023winert}, the authors explored a neural ray-tracing-based construction framework, where neural networks (NNs) are employed to iteratively learn ray-scene interactions.
Building upon this, a generalizable neural ray-tracing method was developed in \cite{Bian2025generalizable}, which facilitates outdoor cross-scenario generalization with improved efficiency.
In addition, the authors in \cite{Lu2024newrf} proposed NeWRF, which incorporates the neural radiance field (NeRF) framework and improves data efficiency by exploiting angle-of-arrival (AoA) as auxiliary input, enabling CKM construction with limited data. 

However, location-based approaches require user locations with wavelength-level accuracy, which is often infeasible to obtain in real-world systems \cite{Chen2025channel}.
Consequently, their practical application remains constrained.
To address this challenge, location-agnostic approaches have gained increasing attention \cite{Liu2021fire,Zhao2023nerf2,Wen2025wrf}, which substitute location inputs in CKMs with data types that are more readily available in current systems.
For instance, the authors in \cite{Liu2021fire} proposed a framework that predicts downlink CSI directly from uplink CSI, which builds upon a variational autoencoder architecture.
Moreover, a NeRF-inspired scheme called NeRF$^2$ was introduced in \cite{Zhao2023nerf2} for CKM construction.
This scheme leverages uplink CSI for downlink prediction while accounting for the propagation characteristics of radio frequency (RF) signals.
Additionally, the authors in \cite{Wen2025wrf} presented WRF-GS, a method based on Gaussian splatting that enables efficient CSI prediction from uplink data with improved computational performance.

Although the aforementioned methods have exhibited competitive accuracy without location information, they typically focused on quasi-static environments, resulting in performance degradation in highly dynamic scenes.
The underlying cause is that the uplink-to-downlink mapping becomes increasingly ill-posed once the temporal dimension is introduced.
A practical solution is to integrate partial downlink pilots, thereby anchoring the mapping while preserving low pilot overhead.
Moreover, existing methods treat uplink CSI inputs as generic tensors and overlook their spatial-spectral characteristics in MIMO-orthogonal frequency division multiplexing (OFDM) systems.
Consequently, the correlations between different antenna elements and carriers have not been adequately captured, leading to compromised performance.
Therefore, it is worth developing an NN that is tailored for these correlations.

In this paper, we propose a scheme named location-agnostic dynamic CKM (LAD-CKM), and our main contributions are summarized as follows.
\begin{itemize}
\item We propose LAD-CKM, a novel CKM framework constructed through dynamic RF radiance field rendering, which is agnostic of user locations and adaptive to environment dynamics, offering an promising solution for practical outdoor dynamic scenarios.
\item We develop a \textbf{Ra}diator \textbf{Re}presentation \textbf{N}etwork (RARE-Net) for RF radiance field rendering in MIMO-OFDM systems, which couples a spatially-aware backbone with dedicated frequency-domain attention to jointly exploit spatial-spectral correlations.
\item To address the challenges posed by dynamic scenes, we propose to use both uplink CSI and partial downlink CSI as the inputs for CKM.
A novel adaptive deformation module is designed to deform the queries of the RARE-Net, enabling adaptiveness to environment dynamics. 
\item Simulation results demonstrate that our LAD-CKM yields significant performance gains compared with existing baselines in terms of effective data rate.
\end{itemize}

The remainder of the paper is organized as follows.
Section \ref{systemModel} describes the system model underlying the LAD-CKM framework.
The detailed designs of LAD-CKM are presented in Section \ref{methods}.
Then, simulation results are provided in Section \ref{Simulation}, and Section \ref{Conclusion} concludes the paper.

\section{System Model} \label{systemModel}
We focus on a MIMO-OFDM frequency-division duplex (FDD) dynamic communication scenario, where the BS is fixed with $N_t$ antennas, the mobile UE has $N_r$ antennas, and the system operates over $N_c$ OFDM subcarriers.

\subsection{Wireless Channel Model} \label{wirelessChModel}
We model the wireless channel based on RF radiance field, which can effectively capture the RF signal distribution in the surrounding environment \cite{Zhao2023nerf2, Lu2024newrf}.
Notably, we innovatively formulate a dynamic RF radiance field for better modeling the wireless channel in dynamic scenes, which is parameterized with respect to the coherence block index $t$.

\begin{figure}[t]
	\centering
	\includegraphics[width=0.45\textwidth]{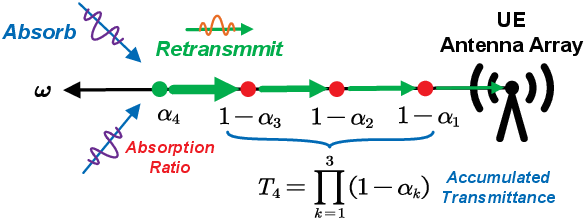}
	\captionsetup{font=footnotesize}
	\caption{Propagation modeling for the radiated signal from a wireless radiator.}
	\label{synthesize}
\end{figure}

For clarity, the basic RF radiance field for a single-antenna, single-carrier system is first analyzed.
To begin with, the environment should first be discretized into a set of volume blocks.
According to the Huygens-Fresnel principle \cite{Zhao2023nerf2}, when the original signal $x(t)$ transmitted by the BS within coherence block $t$ reaches a volume block through multiple propagation paths, this block can be treated as a secondary source that re-transmits $x(t)$.
Based on this principle, each volume block is modeled as a virtual radiator, which absorbs signals from all directions and re-transmits them with an absorption ratio $\alpha(t)$, which is determined by the radiator's instantaneous electronic-magnetic (EM) property $\sigma(t)$ and block length $\delta$, given by
\begin{IEEEeqnarray}{rCl}
	\alpha(t)=1-e^{-\sigma(t)\delta}.
\end{IEEEeqnarray}
Furthermore, the signals re-transmitted by any radiator are again partially absorbed by the obstructing radiators that lie between it and the UE.
This successive effect is modeled as accumulated transmittance $T(t)$, given as
\begin{IEEEeqnarray}{rCl}
	T_{j}(t)=\prod_{k=1}^{j-1}(1-\alpha_{k}(t)),
\end{IEEEeqnarray}
where $j$ and $k$ indicate the radiator index along a propagation path towards the UE.
The propagation of the radiated signal from a virtual radiator is illustrated in Fig. \ref{synthesize}.

To predict the downlink CSI, a set of rays $\{\bm{\omega}_i\}_{i=1}^{N_a}$ originating from the UE's antenna array should be sampled first, where $N_a$ denotes the number of rays across the angular domain.
Along each ray, a sequence of virtual radiators is sampled at intervals $\{\delta_{ij}\}_{j=1}^{N_s}$, where $i$ indicates the ray index and $N_s$ represents the number of sampled radiators along the radial direction.
With the radiators sampled, the corresponding radiated signal and EM property of each radiator can be obtained from an NN $\mathcal{G}_{\bm{\Theta}}(\cdot)$, where $\bm{\Theta}$ denotes the NN's parameter set.
This NN takes as inputs the uplink CSI $h^{\mathrm{U}}(t)$ at each radiator and its radiance direction $\bm{\omega}(t)$, and outputs the corresponding radiated signal $s(t)$ and EM property $\sigma(t)$, given as
\begin{IEEEeqnarray}{rCl}
	\mathcal{G}_{\bm{\Theta}}(\cdot): (h^\mathrm{U}(t),\bm{\omega}(t))\longrightarrow(s(t),\sigma(t)).\label{basicFieldMapping}
\end{IEEEeqnarray}
Subsequently, the received signal at the UE is computed as
\begin{IEEEeqnarray}{rCl}
	\hat{y}(t) = \sum_{i=1}^{N_a}\sum_{j=1}^{N_s}T_{ij}(t)\alpha_{ij}(t)s_{ij}(t).
\end{IEEEeqnarray}
Finally, the downlink CSI can be expressed as
\begin{IEEEeqnarray}{rCl}
	\hat{h}^\mathrm{D}(t) = \frac{\hat{y}(t)}{x(t)} = \sum_{i=1}^{N_a}\sum_{j=1}^{N_s}T_{ij}(t)\alpha_{ij}(t)c_{ij}(t),\label{basicFieldChannel}
\end{IEEEeqnarray}
where $c_{ij}(t) = s_{ij}(t)/x(t)$ represents the effect of aggregating all the signals arriving at the virtual radiator.
Since we are interested in the downlink CSI rather than the received signal, the radiated signal $s(t)$ in the NN's outputs can be changed to the radiator aggregating coefficient $c(t)$.

\begin{figure*}[t]
	\centering
	\includegraphics[width=0.99\textwidth]{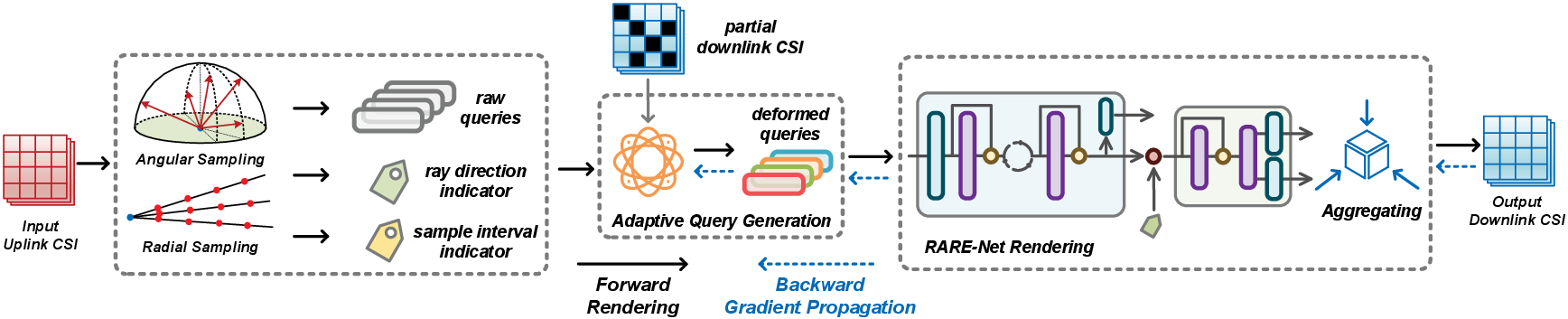}
	\captionsetup{font=footnotesize}
	\caption{The construction pipeline of LAD-CKM.}
	\label{pipeline}
\end{figure*}

Building upon the basic formulation, we commence by delving into the formulation of MIMO-OFDM RF radiance fields.
We denote the transmitted and received signals with subcarrier frequency $f$ as $\mathbf{x}(f,t)\in\mathbb{C}^{N_t}$ and $\mathbf{y}(f,t)\in\mathbb{C}^{N_r}$, respectively.
To reveal the interactions between individual antenna elements, we model each radiator as a cluster of $N_r\times N_t$ sub-radiators, where each sub-radiator corresponds to a unique transmit-receive antenna pair.
Hence, the EM property of the $j$-th radiator in the $i$-th ray is represented as a matrix $\bm{\sigma}_{ij}(f,t)\in\mathbb{R}^{N_r\times N_t}$, composed of the individual sub-radiators’ EM properties.
Similarly, the absorption ratio and accumulated transmittance are represented as $\bm{\alpha}_{ij}(f,t)\in\mathbb{R}^{N_r\times N_t}$ and $\mathbf{T}_{ij}(f,t)\in\mathbb{R}^{N_r\times N_t}$, respectively.
Furthermore, the radiated signal is denoted as $\mathbf{s}_{ij}(f,t)\in\mathbb{C}^{N_r}$.
Then, the radiator aggregating coefficients can be derived as a matrix $\mathbf{C}_{ij}(f,t)\in\mathbb{C}^{N_r\times N_t}$, given by
\begin{IEEEeqnarray}{rCl}
	\mathbf{C}_{ij}(f,t) =\frac{1}{P} \mathbf{s}_{ij}(f,t)\mathbf{x}^H(f,t),
\end{IEEEeqnarray}
where $P=\mathbf{x}^H(f,t)\mathbf{x}(f,t)$ denotes the transmit power of the BS.
Finally, Eq. (\ref{basicFieldChannel}) can be rewritten as
\begin{IEEEeqnarray}{rCl}
	\hat{\mathbf{H}}^\mathrm{D}(f,t) = \sum_{i=1}^{N_a}\sum_{j=1}^{N_s}\bm{\alpha}_{ij}(f,t)\odot\mathbf{T}_{ij}(f,t)\odot\mathbf{C}_{ij}(f,t),\label{finalFormulation}
\end{IEEEeqnarray}
where $\odot$ denotes element-wise multiplication operators for matrices.
Additionally, the NN function Eq. (\ref{basicFieldMapping}) is updated as
\begin{IEEEeqnarray}{rCl}
	\mathcal{G}_{\bm{\Theta}}(\cdot): (\mathbf{H}^\mathrm{U}(f,t),\bm{\omega}(f,t))\longrightarrow(\mathbf{C}(f,t),\bm{\sigma}(f,t)).\label{finalMapping}
\end{IEEEeqnarray}
\subsection{Performance Metric} \label{metric}
For each subcarrier $f$ within coherence block $t$, the UE receives
\begin{IEEEeqnarray}{rCl}
	\mathbf{y}(f,t) = \mathbf{w}^H(t)\mathbf{H}^{\mathrm{D}}(f,t)\mathbf{f}(t)\mathbf{x}(f,t)+\mathbf{w}^H(t)\mathbf{n}(f,t),
\end{IEEEeqnarray}
where $\mathbf{n}(f,t)\in\mathbb{C}^{N_r}$ is the additive white Gaussian noise (AWGN) with the distribution $\mathcal{CN}(0, \sigma^2\mathbf{I})$, $\mathbf{f}(t)\in\mathbb{C}^{N_t}$ and $\mathbf{w}(t)\in\mathbb{C}^{N_r}$ denote the transmit and receive beamforming vectors, respectively, which are chosen to maximize the channel gain based on the predicted downlink CSI $\hat{\mathbf{H}}^\mathrm{D}(f,t)$, given by
\begin{equation}
	(\mathbf{w}_*(t),\mathbf{f}_*(t)) = \arg \max_{\mathbf{f}\in\mathcal{F},\mathbf{w}\in\mathcal{W}} \frac{1}{N_c}\sum_{f}||\mathbf{w}^H\hat{\mathbf{H}}^{\mathrm{D}}(f,t)\mathbf{f}||^2,
\end{equation}
where $\mathcal{F}$ and $\mathcal{W}$ are the codebooks at the BS and UE, respectively.
The corresponding effective data rate is computed as
\begin{equation}
	R(t) = \frac{1-\rho}{N_c}\sum_f\log_2\Bigl(1+\gamma\cdot||\mathbf{w}^H_*(t)\mathbf{H}^{\mathrm{D}}(f,t)\mathbf{f}_*(t)||^2\Bigr),
\end{equation}
where $\gamma=P/\sigma^2$ denotes the transmit signal-to-noise-ratio (SNR) and $\rho$ indicates the pilot overhead ratio of space-time-frequency resources.
We select $R(t)$ as the key metric for evaluating the effectiveness of LAD-CKM.
\section{Construction of LAD-CKM} \label{methods}
In this section, we provide an in-depth elaboration on the proposed LAD-CKM and its implementation specifics.
\subsection{Construction Pipeline}
The overall construction pipeline of LAD-CKM is illustrated in Fig. \ref{pipeline}.
The normalized mean squared error (NMSE) loss $\mathcal{L}_{\mathrm{NMSE}}$ is adopted for NN optimization, given as
\begin{equation}
	\mathcal{L}_{\mathrm{NMSE}} = \frac{1}{NN_bN_c}\sum_{i=1}^{N}\sum_{t}\sum_{f}{\frac{||\mathbf{\hat{H}}^\mathrm{D}_{i}(f,t)-\mathbf{H}^\mathrm{D}_{i}(f,t)||^2}{||\mathbf{H}^\mathrm{D}_{i}(f,t)||^2}},
\end{equation}
where $\mathbf{\hat{H}}^\mathrm{D}_{i}(f,t)$ and $\mathbf{H}^\mathrm{D}_{i}(f,t)$ represent the predicted and ground truth CSI of subcarrier frequency $f$ within coherence block $t$ in the $i$-th sample, respectively, with $N$ and $N_b$ denoting the total number of data samples and coherence blocks, respectively.
In particular, this pipeline comprises three major stages:
\begin{itemize}
	\item \textit{Virtual Radiator Sampling:}
	This stage aims to effectively sample the virtual radiators within the scene, which involves angular and radial sampling.
	\item \textit{Adaptive Query Generation:}
	This stage generates queries for the RARE-Net by preprocessing the raw inputs.
	It leverages two geometrical indicators to provide sampling information and performs adaptive query-deformation based on partial instantaneous downlink CSI observations, which enables adaptation to environment dynamics.
	\item \textit{RARE-Net Rendering:}
	This stage queries the RARE-Net to render the radiators, and finally outputs the downlink CSI prediction by radiator aggregation.
\end{itemize}

\subsection{Virtual Radiator Sampling}
Generally, there exists an inherent trade-off between sampling resolution and computational complexity.
While fine-grained sampling can naturally improve accuracy, it also increases computational overhead.
To achieve accurate yet low-complexity CKM construction, it is crucial to adopt an effective sampling strategy within a given sampling resolution.
Specifically, we employ the spherical Fibonacci grid (SFG) algorithm \cite{Swinbank2006fibonacci} for angular sampling while employing uniform sampling along the radial dimension.

\subsection{Adaptive Query Generation}

After radiator sampling, we obtain $N_s$ radiators spaced at intervals $\delta_{ij}(j=1,...,N_s)$ for each ray $\bm{\omega}_i(i=1,...,N_a)$ originating from the UE's antenna array.
To train the RARE-Net, the corresponding uplink CSI at each sampled radiator, $\mathbf{H}^\mathrm{U}_{ij}\in\mathbb{R}^{2N_c\times N_r \times N_t}$, is required as queries\footnote{Complex CSI is converted to real by concatenating real and imaginary parts along the frequency dimension. To facilitate elaboration, frequency and time indices $(f,t)$ are omitted unless necessary.}.
However, only the uplink CSI at the UE's antenna array  $\mathbf{H}^\mathrm{U}_{\mathrm{RX}}\in\mathbb{R}^{2N_c\times N_r \times N_t}$ can be obtained, and deriving a closed-form mapping from $\mathbf{H}^\mathrm{U}_{\mathrm{RX}}$ to each $\mathbf{H}^\mathrm{U}_{ij}$ is analytically intractable.
To address this challenge, we aim to approximate $\mathbf{H}^\mathrm{U}_{ij}$ based on $\mathbf{H}^\mathrm{U}_{\mathrm{RX}}$ via deep learning techniques.

\begin{figure}[t]
	\centering
	\includegraphics[width=0.45\textwidth]{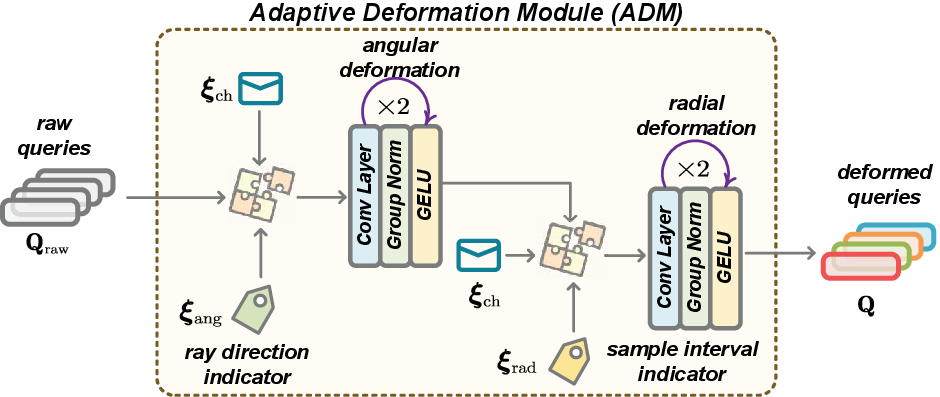}
	\captionsetup{font=footnotesize}
	\caption{The architecture of the ADM.}
	\label{shapingFilter}
\end{figure}

Specifically, we first perform a straight-through approximation by directly replicating $\mathbf{H}^\mathrm{U}_{\mathrm{RX}}$ to each radiator, composing raw queries to the RARE-Net, denoted as $\mathbf{Q}_{\mathrm{raw}}\in\mathbb{R}^{N_a\times N_s\times 2N_c\times N_r \times N_t}$.
Nevertheless, this coarse approximation overlooks the uplink CSI variations caused by the spatial distributions of the radiators, potentially leading to biased rendering of the RF radiance field and inefficient learning.
To further refine this approximation, we introduce a query deformation process by developing an adaptive deformation module (ADM).
The ADM learns to fine-tune $\mathbf{Q}_{\mathrm{raw}}$ on its angular and radial dimensions by jointly considering the partially observed downlink CSI and the geometrical distribution of the radiators, thereby achieving a more accurate approximation.
Notably, the geometrical distribution of the radiators is determined by the ray directions $\bm{\omega}\in\mathbb{R}^{N_a\times 3}$ and sampling intervals $\bm{\delta}\in\mathbb{R}^{N_a\times N_s}$.
Inspired by this, we design the ADM as a hierarchical module, which operates in two stages with the guidance of these two indicators, as depicted in Fig. \ref{shapingFilter}.

During the first stage, the ADM fine-tunes $\mathbf{Q}_{\mathrm{raw}}$ along its angular dimension.
Specifically, the ray direction indicator $\bm{\omega}$ is first expanded to size $N_a\times N_s\times 3\times N_r \times N_t$ by replicating, and then reshaped into size $N_a N_s\times 3\times N_r \times N_t$, denoted as
\begin{IEEEeqnarray}{rCl}
	\bm{\xi}_{\mathrm{ang}}=\mathcal{R}(\bm{\omega})\in\mathbb{R}^{N_a N_s\times 3\times N_r \times N_t}.
\end{IEEEeqnarray}
Meanwhile, the partial downlink CSI $\mathbf{\hat{H}}^{\mathrm{D}}_{\mathrm{part}}$ is first zero-padded to size $2N_c\times N_r\times N_t$, then expanded to size $N_a\times N_s\times 2N_c\times N_r \times N_t$ by replicating, and finally reshaped into size $N_a N_s\times 2N_c\times N_r \times N_t$, denoted as
\begin{IEEEeqnarray}{rCl}
	\bm{\xi}_{\mathrm{ch}}=\mathcal{R}(\mathbf{\hat{H}}^{\mathrm{D}}_{\mathrm{part}})\in\mathbb{R}^{N_a N_s\times 2N_c\times N_r \times N_t}.
\end{IEEEeqnarray}
Subsequently, $\mathbf{Q}_{\mathrm{raw}}$ is reshaped into size $N_a N_s\times 2N_c\times N_r \times N_t$, and concatenated with the indicators along the channel dimension
\begin{IEEEeqnarray}{rLl}
	 \mathbf{Q}_{\mathrm{raw}}\oplus_2\bm{\xi}_{\mathrm{ch}}\oplus_2\bm{\xi}_{\mathrm{ang}}\in\mathbb{R}^{N_a N_s\times (4N_c+3)\times N_r \times N_t},
\end{IEEEeqnarray}
where $\oplus_2$ represents the tensor concatenation operation along the second dimension.
Next, the concatenated tensor is processed by the angular deformation layer $\mathcal{M}_{\bm{\theta}}(\cdot)$
\begin{equation}
	\mathbf{Q}_{\mathrm{ang}}=\mathcal{M}_{\bm{\theta}}(\mathbf{Q}_{\mathrm{raw}}\oplus_2\bm{\xi}_{\mathrm{ch}}\oplus_2\bm{\xi}_{\mathrm{ang}})\in\mathbb{R}^{N_a N_s\times 2N_c\times N_r \times N_t},
\end{equation}
where $\mathbf{Q}_{\mathrm{ang}}$ represents the queries fine-tuned on the angular dimension and $\bm{\theta}$ denotes the trainable parameters of the angular deformation layer.

\begin{figure}[t]
	\centering
	\includegraphics[width=0.5\textwidth]{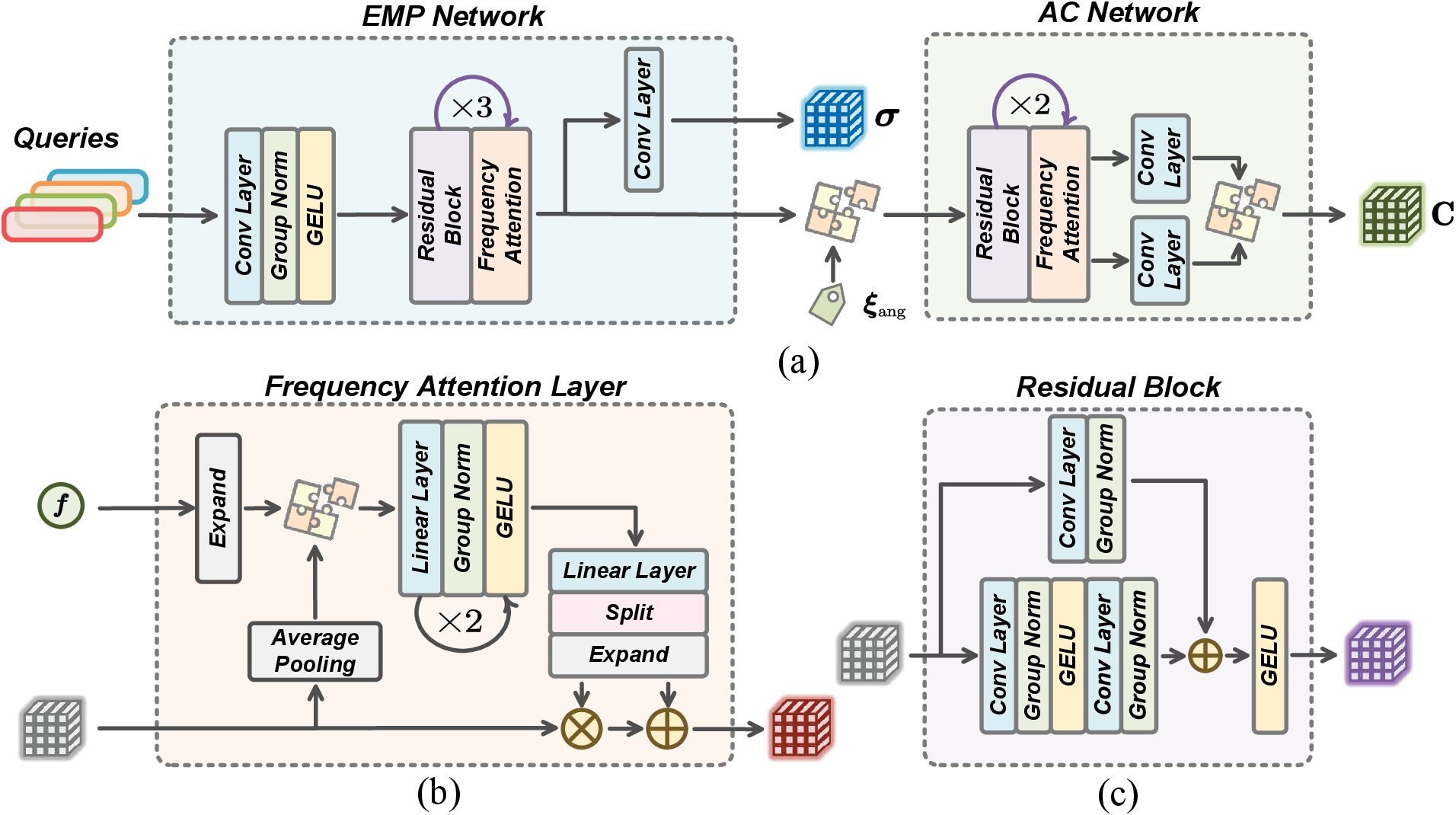}
	\captionsetup{font=footnotesize}
	\caption{The architecture of the RARE-Net. (a) provides an overview of RARE-Net's entire architecture. (b) and (c) depict the inner structures of the frequency-attention layer and residual blocks within the RARE-Net, respectively.}
	\label{RARE-Net}
\end{figure}

In the second stage, the ADM refines $\mathbf{Q}_{\mathrm{ang}}$ along its radial dimension.
Similar to the first stage, the sampling interval indicator $\bm{\delta}$ is first expanded to size $N_a\times N_s\times 2N_c\times N_r \times N_t$ by replicating, and then reshaped into size $N_a \times N_s\times 2N_c\times N_r N_t$, denoted as
\begin{IEEEeqnarray}{rCl}
	\bm{\xi}_{\mathrm{rad}}=\mathcal{R}(\bm{\delta})\in\mathbb{R}^{N_a\times N_s\times 2N_c\times N_r N_t}.
\end{IEEEeqnarray}
Following this, $\mathbf{Q}_{\mathrm{ang}}$ and $\bm{\xi}_{\mathrm{ch}}$ are reshaped into size $N_a \times N_s\times 2N_c\times N_r N_t$, and concatenated with $\bm{\xi}_{\mathrm{rad}}$ along the channel dimension
\begin{IEEEeqnarray}{rLl}
	\mathbf{Q}_{\mathrm{ang}}\oplus_2\bm{\xi}_{\mathrm{ch}}\oplus_2\bm{\xi}_{\mathrm{rad}}\in\mathbb{R}^{N_a\times 3N_s\times 2N_c\times N_rN_t}.
\end{IEEEeqnarray}
Subsequently, the concatenated tensor is passed through the radial deformation layer $\mathcal{M}_{\bm{\phi}}(\cdot)$, given as
\begin{align}
	\mathbf{Q}_{\mathrm{rad}} & =  \mathcal{M}_{\bm{\phi}}(\mathbf{Q}_{\mathrm{ang}}\oplus_2\bm{\xi}_{\mathrm{ch}}\oplus_2\bm{\xi}_{\mathrm{rad}})\in\mathbb{R}^{N_a\times N_s\times 2N_c\times N_r N_t},\\
	\mathbf{Q} & =  \mathcal{R}(\mathbf{Q}_{\mathrm{rad}})\in\mathbb{R}^{N_aN_s\times 2N_c\times N_r \times N_t},
\end{align}
where $\mathbf{Q}_{\mathrm{rad}}$ denotes the queries fine-tuned along the radial dimension, $\bm{\phi}$ is the trainable parameter set of the radial deformation layer, and $\mathbf{Q}$ represents the final output of the ADM.

In summary, the entire query generation process can be concisely represented by the following expression
\begin{IEEEeqnarray}{rCl}
	\mathbf{Q} = \mathcal{M}_{\bm{\phi}}\bigl(\mathcal{M}_{\bm{\theta}}(\mathbf{Q}_{\mathrm{raw}}\oplus_2\bm{\xi}_{\mathrm{ch}}\oplus_2\bm{\xi}_{\mathrm{ang}})\oplus_2\bm{\xi}_{\mathrm{ch}}\oplus_2\bm{\xi}_{\mathrm{rad}}\bigr),\IEEEeqnarraynumspace
\end{IEEEeqnarray}
where the reshaping operation is implicitly incorporated.
The trainable parameters $\{\bm{\theta},\bm{\phi}\}$ of the ADM are jointly optimized with the RARE-Net in an end-to-end manner, as described in the following subsection.

\subsection{RARE-Net Rendering}

The RARE-Net is designed to predict the EM properties and aggregating coefficients of all the sampled virtual radiators, thus rendering the entire RF radiance field.
The network architecture is presented in Fig. \ref{RARE-Net}.
\subsubsection{Spatial-Aware Backbone}
Existing methods \cite{Zhao2023nerf2,Lu2024newrf,Wen2025wrf} typically implement their models based on fully-connected layers.
These simple structures are insufficient for effectively capturing the spatial correlations within CSI data, thereby limiting the networks' capacity in multiple antenna systems.
In contrast, the RARE-Net incorporates 2D convolutional layers. This design significantly enhances the network’s capability to model spatial correlations in CSI.

Specifically, RARE-Net consists of two sub-networks: the EM property (EMP) network and the aggregating coefficient (AC) network, tasked with predicting the EM properties and aggregating coefficients, respectively.
The input $\mathbf{Q}$ is initially processed by the EMP network, which contains an up-sampling convolutional block and three residual blocks, the inner structure of the latter is depicted in Fig. \ref{RARE-Net}(c).
The output features from these blocks then diverge into two distinct processing flows.
The first flow involves a down-sampling convolutional layer that produces the prediction of EM properties $\bm{\sigma}$.
The second flow directs the features into the AC network for predicting the aggregating coefficients $\mathbf{C}$.
Importantly, we model the virtual radiators as anisotropic sources that radiate direction-dependent signals, which enhances the network’s ability to capture multipath propagation effects.
To facilitate the learning of these directional characteristics, the output features from the EMP network are concatenated with the ray direction indicator $\bm{\xi}_{\mathrm{ang}}$ before being fed into the AC network.
Regarding the AC network, the input features first pass through a down-sampling residual block, succeeded by an up-sampling residual block and a down-sampling convolutional layer.
Finally, the predicted aggregating coefficients $\mathbf{C}$ is generated in a real-valued format, with its real and imaginary parts concatenated along the frequency dimension.
The predicted $\bm{\sigma}$ and $\mathbf{C}$ are then leveraged to obtain the final prediction of the downlink CSI $\hat{\mathbf{H}}^\mathrm{D}$ based on Eq. (\ref{finalFormulation}).

\subsubsection{Frequency-Aware Fine-Tuning}

To further enhance RARE-Net's prediction performance beyond its spatial-aware backbone design, we introduce frequency-aware fine-tuning by developing the frequency-attention layer.
This layer serves as a feature modulator that refines intermediate features by incorporating explicit frequency values of each subcarrier $f$ (GHz) as side information.

Specifically, the frequency-attention layer fine-tunes the input features via affine transformations.
The weights and biases for these transformations are jointly determined by both the input features and the frequency values of each subcarrier.
As presented in Fig. \ref{RARE-Net}(b), the input features and frequency values are first processed by average pooling and dimension expansion, respectively, to align their dimensions and enable the merging of these two information streams.
Subsequently, three fully-connected layers, combined with group normalization layers and GELU activations, are employed to process the merged information.
Finally, the weights and biases for the final affine transformation on the input features are generated from the processed information.
It is important to note that both $\bm{\sigma}$ and $\mathbf{C}$ are frequency-dependent.
Therefore, the frequency-attention layers are embedded into both sub-networks, attached to the tail of the residual blocks, as illustrated in Fig. \ref{RARE-Net}(a).

\section{Simulation Results} \label{Simulation}

\begin{figure}[t]
	\centering
	\includegraphics[width=0.4\textwidth]{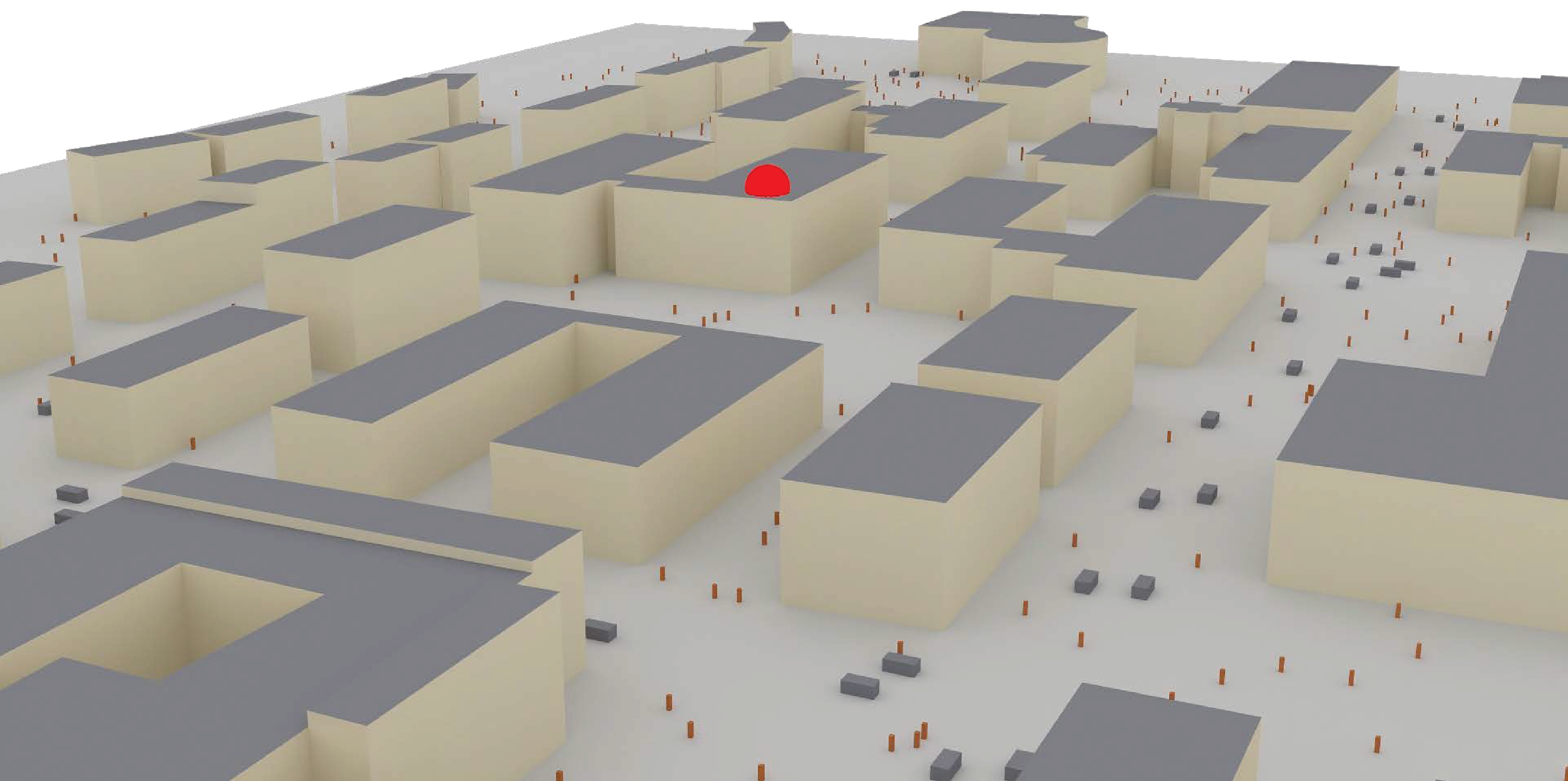}
	\captionsetup{font=footnotesize}
	\caption{The dynamic campus scene for channel dataset generation.}
	\label{campus_scene}
\end{figure}

To evaluate the proposed methods, a novel synthetic channel dataset is created in outdoor dynamic scenes via ray-tracing.
As shown in Fig.~\ref{campus_scene}, we consider a real-world campus environment whose 3D geometry is downloaded from OpenStreetMap\footnote{https://www.openstreetmap.org/}.
The scene is imported into Blender for material parameterization and then fed into the Sionna ray-tracing platform \cite{Hoydis2023sionna} to generate ground-truth MIMO-OFDM channels.
To simulate environment dynamics, we randomly reposition dynamic scatterers in every coherence block, where pedestrians are modeled as dry dielectric boxes ($0.5\,\text{m}\times 0.5\,\text{m}\times 1.8\,\text{m}$) and vehicles as metal boxes ($2\,\text{m}\times 4\,\text{m}\times 1.6\,\text{m}$).
The OFDM bandwidth is $20$ MHz, split into $64$ sub-carriers with $312.5$ kHz spacing, where $52$ sub-carriers are used for channel measurement.
The central frequencies are set to $6.715$ GHz (uplink) and $6.765$ GHz (downlink).
Both BS and UE employ uniform planar arrays (UPAs) with discrete Fourier transform (DFT) codebooks.
We consider BS antenna counts $N_t\in\{8,16,32,64,128\}$ and UE antenna counts $N_r\in\{1,4\}$.
For each MIMO configuration $(N_t,N_r)$, $20$ coherence blocks are generated, with $2,500$ UEs randomly dropped per block.

We implement our LAD-CKM and all baselines using the PyTorch platform, with a $2.60$ GHz Intel(R) Xeon(R) Platinum 8350C CPU, and an NVIDIA GeForce RTX 4090 GPU.
To train LAD-CKM, the Adam optimizer with an initial learning rate of $5\times 10^{-5}$ and a batch size of $2$ is employed.
The learning rate is adjusted over time using the ReduceLROnPlateau scheduler with a patience of $10$ and a reduction factor of $0.9$.
We compare the effective data rate performance of the following schemes:
\begin{itemize}
	\item Perfect CSI: The upper bound, which assumes the downlink CSI is perfectly known.
	\item LAD-CKM ($\rho=0.125$): The proposed scheme with $12.5\%$ pilot overhead, where pilots are transmitted on every-$4$th subcarrier and every-$2$nd antenna.
	\item LAD-CKM ($\rho=0$): A purely location-agnostic baseline that relies solely on uplink CSI and transmits no downlink pilots.
	\item NeRF$^2$ \cite{Zhao2023nerf2}: A NeRF-based approach that
	predicts downlink CSI based on uplink CSI.
	\item FIRE \cite{Liu2021fire}: A variational autoencoder-based method that predicts downlink CSI from uplink CSI.
	We utilize the open-source code available for this method.
\end{itemize}

\begin{figure}[t]
	\centering
	\includegraphics[width=0.45\textwidth]{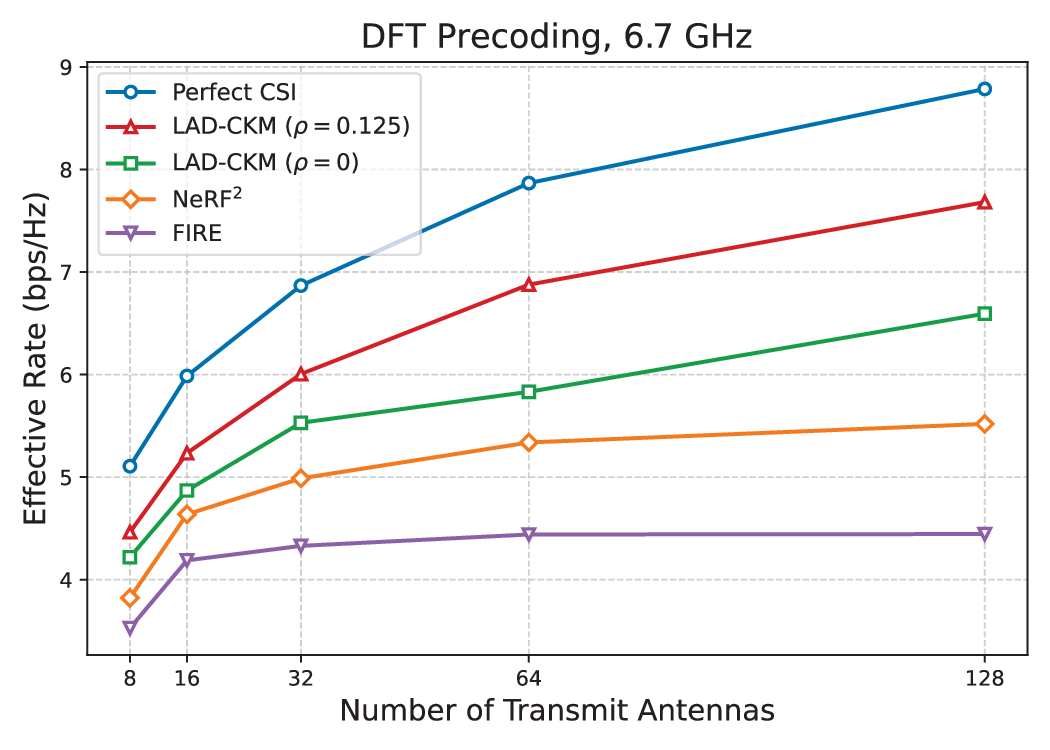}
	\captionsetup{font=footnotesize}
	\caption{The performance of different schemes versus antenna numbers.}
	\label{sim_nta}
\end{figure}

Fig. \ref{sim_nta} compares effective data rate performance under different MIMO configurations, specifically $N_r=1$ and $N_t\in\{8, 16, 32, 64, 128\}$.
It is readily seen that LAD-CKM consistently outperforms all baselines, especially when $N_t$ is large.
This superiority is attributed to the ADM design, which adapts LAD-CKM to environment dynamics by query deformation mechanisms.
Moreover, even with pilots removed ($\rho=0$), the performance of LAD-CKM remains higher than other benchmarks, confirming that the RARE-Net effectively exploits spatial-spectral correlations within wireless channel. 

\begin{figure}[t]
	\centering
	\includegraphics[width=0.45\textwidth]{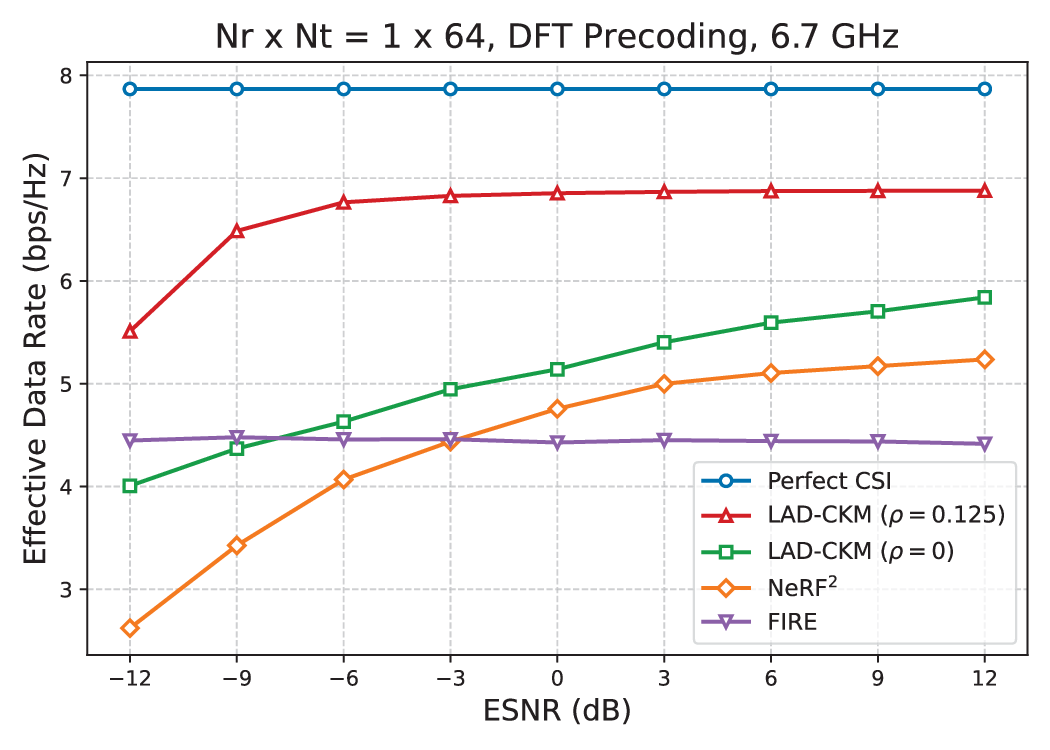}
	\captionsetup{font=footnotesize}
	\caption{The performance of different schemes versus ESNR.}
	\label{sim_cea}
\end{figure}

Fig. \ref{sim_cea} evaluates effective data rate performance under imperfect CSI inputs.
To simulate these imperfections, both uplink and partial downlink CSI estimates are corrupted with AWGN, yielding estimation signal-to-noise ratio (ESNR) values ranging from $-12$ dB to $12$ dB, where
\begin{IEEEeqnarray}{rCl}
	\mathrm{ESNR} = 10\log_{10}(\frac{||\mathbf{H}||^2}{||\mathbf{\hat{H}}-\mathbf{H}||^2}),
\end{IEEEeqnarray}
$\mathbf{H}$ and $\hat{\mathbf{H}}$ represent the ground-truth and estimated CSI, respectively.
All schemes are trained with perfect inputs and evaluated without retraining.
It is clear that LAD-CKM maintains robust to estimation errors across the entire ESNR range, highlighting its reliability for practical applications.

\section{Conclusion} \label{Conclusion}

In this paper, we introduced LAD-CKM, an innovative CKM framework agnostic of user locations and adaptive to environment dynamics.
Specifically, we proposed a dynamic radiance field rendering-based approach, which takes instantaneous uplink CSI and partial downlink CSI as inputs.
To enable effective rendering, we designed a dedicated RARE-Net to capture the spatial-spectral correlations within the inputs.
Furthermore, we developed ADM, a module devised to deform the queries of RARE-Net according to instantaneous channel dynamics, thereby enhancing CSI prediction accuracy under mobility.
Simulation results demonstrate that LAD-CKM yields significant performance gains compared with existing baselines in terms of effective data rate.

\bibliographystyle{IEEEtran}
\bibliography{IEEEabrv,Reference}

\end{document}